\documentstyle[aps,preprint]{revtex}
\begin{document}
\tightenlines

\title{Theory of the NO+CO Surface Reaction Model}
\author{Adriana G. Dickman$^{a}$,
Bartira C. S. Grandi$^{b}$,
Wagner Figueiredo$^{c}$, and
Ronald Dickman$^{d,e}$}
\address{Departamento de F\'\i sica,
Universidade Federal de Santa Catarina, Campus Universit\'ario - Trindade,
CEP 88040-900, Florian\'opolis - SC, Brasil}
\date{\today}
\maketitle
\begin{abstract}
We derive a pair approximation (PA) for the NO+CO
model with instantaneous reactions.
For both the triangular and square lattices, the PA,
derived here using a simpler approach, yields
a phase diagram with an active state for CO-fractions $y$
in the interval $y_1 < y < y_2$, with a continuous
(discontinuous) phase transition to a poisoned state
at $y_1$ ($y_2$).  This is in qualitative 
agreement with simulation for the triangular lattice,
where our theory gives
a rather accurate prediction for $y_2$.
To obtain the correct phase diagram
for the square lattice, i.e., {\it no} active stationary state,
we reformulate the PA using {\it sublattices}.
The (formerly) active regime is then replaced by a poisoned state
with broken symmetry (unequal sublattice coverages), as observed
recently by Kortl\"uke et al. [Chem. Phys. Lett. {\bf 275}, 85 (1997)].
In contrast with their approach, in which the active state persists,
although reduced in extent, we report here the first qualitatively
correct theory of the NO+CO model on the square lattice.
Surface diffusion of nitrogen can lead to an active state
in this case.
In one dimension, the PA predicts that diffusion
is required for the existence of an active state.
\vspace {0.3truecm}

\noindent PACS numbers: 05.70.Ln, 82.65.Jv, 82.20.Mj, 05.70.Fh 
\end{abstract}
\vspace{1.0truecm}

\noindent $^a${\small electronic address: dri@fisica.ufsc.br } \\
$^b${\small electronic address: bartirag@fisica.ufsc.br } \\
$^c${\small electronic address: wagner@fisica.ufsc.br } \\
$^d${\small electronic address: dickman@fisica.ufsc.br } \\
$^e${\small On leave of absence from: Department of Physics and Astronomy,
Herbert H. Lehman College, City University of New York,
Bronx, NY, 10468-1589.} \\

\newpage

\section{Introduction}

Following the introduction by Ziff, Gulari, and Barshad 
of a simple lattice model (ZGB) for the kinetics of the reaction
CO + 1/2 O$_2$ $\rightarrow$ CO$_2$ on a catalytic surface,
the study of surface reaction models has attracted increasing
attention in nonequilibrium statistical physics \cite{zgb}.
Motivated by possible applications as well as intrinsic
interest, the phase diagrams of a wide
variety of models have been investigated in simulations and
approximate, mean-field-like analyses.
A typical feature is the existence
of one or more absorbing states, i.e., configurations
from which the system cannot escape \cite{reviews}.
Continuous phase transitions
to an absorbing state fall generically
in the class of directed percolation \cite{RFT,Torre,glb}.
While this aspect is highly universal,
other details of the phase diagram depend on very specific
model-dependent properties such as
steric or geometric effects, the possibility of non-reactive
desorption, diffusion of, and interactions among, adsorbed species.
Applied to surface reaction models, mean-field theories,
particularly at the two-site or
{\it pair} level, often provide reasonable qualitative
predictions for the phase diagram.

In the present work we derive pair mean-field approximations for
one of the more complicated surface reaction models,
that of NO+CO \cite{yk91}.  We briefly review its main features,
deferring a precise definition to Sec. II.
The catalytic surface (i.e., one of the platinum-group metals), is
modelled by a regular lattice (typically square or triangular) of
equivalent adsorption sites.
This surface is exposed to a reservoir of CO and NO at fixed concentrations. 
While CO needs but a single vacant site, NO requires a nearest-neighbor
pair of sites to adsorb.
(We note that a more realistic model, for example of the NO+CO reaction
on Pt(100), would permit NO to adsorb at a single site; a vacant neighbor is
required for dissociation \cite{mwe94}.)
The fundamental control parameter of the model is
$y$, the probability that the next molecule arriving at
the surface will be CO.  One may also introduce nearest-neighbor
hopping rates for the various adsorbed species.
Nearest-neighbor CO-O and N-N pairs are highly reactive: 
if any form (by adsorption or diffusion), they are eliminated
before anything else happens,
and the products (CO$_2$ and N$_2$, respectively)
desorb immediately.

On the triangular lattice (i.e., coordination number six),
the phase diagram of the NO+CO model resembles
that of the ZGB model: 
there is a reactive window for $y_1 < y < y_2$,
with a continuous phase transition at $y_1 \approx 0.17 $, and a discontinuous
transition at $y_2 \approx 0.35$ \cite{yk91,bz,mwe,ij}.  For $y$
values outside the reactive window, the system eventually falls into
an absorbing or ``poisoned" configuration, devoid of vacant sites
(the number of such configurations grows exponentially
with system size).  For $y < y_1$ the final configuration consists
predominantly of O, with an appreciable fraction of N.
(A special case is $y=0$, corresponding to a kind of random sequential
adsorption (RSA) \cite{evans} of dimers, with partial reaction.  The final state
consists of O and N atoms, with isolated vacancies interspersed.)
For $y > y_2$, CO takes over the role played by O in the small-$y$ case.
Diffusion of N, CO, and/or O shifts the transition points to some extent,
but does not modify the phase diagram in any fundamental way.

On the square lattice, the picture is radically different, there being
(without diffusion of N), {\it no} active stationary state, whatever
the value of $y$ \cite{yk91,yk92,yk93,yk93L}.  This observation, based on
Monte Carlo simulations, was explained
by Brosilow and Ziff (BZ), who argued that the active state is unstable to
the filling of one sublattice with N atoms \cite{bz}.  Once this occurs,
further adsorption of dimers is blocked, and the remaining vacancies are
filled in with CO, yielding a poisoned configuration.  While BZ
cast their argument in terms of global sublattice filling, in practice
the system poisons through the growth of local ``antiferromagnetic"
domains, i.e., patches having one or another sublattice 
filled with N.  Since the dynamics stops once the domains fill the system,
relaxation to a globally-ordered state is not possible.
For the same reason, no sharp transitions in coverages are
observed as $y$ is varied \cite{yk91,yk92,yk93,bz,mwe,yk94}.
(It may be overstating the case to say that currently available
results rule out any phase transition in the square lattice.
Simulations show the coverages changing rapidly over a narrow
range of $y$, but without discontinuities in the coverages or their slopes.)
That the N-sublattice instability is responsible for destroying
the active state (despite the absence of global sublattice order), is
indicated by the observations that (1) an active state exists in the
triangular lattice (which does not admit a decomposition into two
sublattices), and (2) that diffusion of N (but not of O,
or of CO) restores the possibility
of an active state \cite{yk94,kn}.  

Several theories of the NO+CO model have been proposed. 
Truncated at the 1-site level, the hierarchy of equations
governing the cluster probabilities yields a
reasonable estimate for $y_2$ on the triangular lattice \cite{bz,mwe},
but places the continuous transition at $y_1 = 0$.
Cort\'es et al. derived a pair approximation for 
the NO+CO model including CO-desorption,
using finite reaction rates \cite{cortes}.
Kortl\"{u}ke, Kuzovkov and von Niessen (KKN)
derived a very accurate prediction for $y_1$ on
the triangular lattice using a two-site cluster approximation \cite{kkn}.

None of the theories mentioned so far gives the phase diagram correctly
for the square lattice: all predict
an active state over some range of $y$.
KKN made the fundamental observation that in this case, one must
allow different concentrations on the two sublattices, to have
any hope of capturing the instability identified by BZ \cite{kkn}.
They devised a two-site cluster approximation
incorporating sublattices, and obtained an active state of
reduced extent, and a {\it third} transition point
inside the (mainly) CO-poisoned phase.  
In other words, the theory of KKN, while representing an improvement
on theories ignoring sublattices, remains at variance with
simulation (and the BZ argument), in allowing an active state,
and predicts a third, unobserved transition.

In this work we formulate a pair approximation (PA) for the NO+CO model
on the square and triangular lattices, as well as in one dimension;
we retain the instantaneous reactions generally used in simulations.
(While a theory employing a finite reaction rate $k$ is simpler algebraically,
its $k \rightarrow \infty $ limit is not
equivalent to a theory with instantaneous reactions \cite{evans91}.)
The standard PA predicts an active state for
$0 < y_1 < y < y_2$; this is qualitatively correct for the
triangular lattice, wrong for the square lattice.
We obtain the correct phase diagram in the latter case from
a pair approximation incorporating sublattices (PAS); the 
regime exhibiting activity in the PA
now poisons via the sublattice instability.
Our results for the coverages are in good accord with
simulation, but due to the assumed homogeneity {\it within}
each sublattice, phase transitions between different kinds
of absorbing states persist at $y_1$ and $y_2$, and the 
order parameter (the difference in sublattice coverages), takes a
nonzero value for $y_1 < y < y_2$.  We find that diffusion of
N atoms lifts the instability, permitting an active state.  But in
our theory, the entire range $y_1 < y < y_2$ becomes active once
the diffusion rate $D_N$ exceeds a critical value.  In simulations the
``active window" opens gradually as $D_N$ is increased \cite{yk94,kn}.

We have devised a simplified approach to deriving cluster
mean-field equations.  The method, which we illustrate
with a simple example, proves particularly useful
in the case of the NO+CO model, which allows eight 
kinds of nearest-neighbor pairs, and up to twenty-two transitions among them.
The remainder of this paper is organized as follows.  Section II
contains a definition of the model, including the reaction and diffusion
steps.  The PA method is described in Sec. III, with examples of
its application to the contact process and the NO+CO reaction given in
Appendices A and B, respectively.  Our results are presented in Sec. IV,
and a brief discussion follows in Sec. V.

\section{Model}

The NO+CO surface reaction model follows the Langmuir-Hinshelwood 
mechanism, in which both reacting species must be adsorbed
on the substrate \cite{oh}. The steps below characterize the model \cite{yk91}:
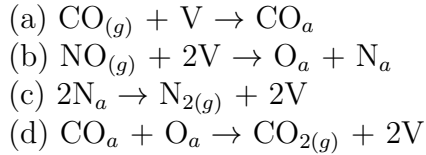
\begin{flushleft}

(a) CO$_{(g)}$ + V  $\rightarrow  $ CO$_{a}$

(b) NO$_{(g)}$ + 2V $\rightarrow $ O$_{a}$ + N$_{a}$

(c) 2N$_{a}          \rightarrow $  N$_{2(g)}$ + 2V 

(d) CO$_{a}$ + O$_{a} \rightarrow $ CO$_{2(g)}$ + 2V \\ 
\end{flushleft}
where A$_a$ indicates an adsorbed species, A$_{(g)}$ a molecule in
the gas phase, and V
a vacant site. Steps (a) and (b) represent adsorption of
carbon monoxide, and of nitric oxide,
respectively. In step (c), two nearest-neighbor nitrogen atoms
combine to form N$_{2(g)}$, and in step (d), an oxygen atom
reacts with a carbon monoxide molecule to form CO$_2$. N$_{2(g)}$
and CO$_{2(g)}$ desorb from the surface immediately. In this work we
assume complete dissociation of NO.  (Various aspects of
incomplete dissociation are considered in
Refs. \cite{yk94,kn,cortes}.)
As noted above, reactions are assumed to occur instantaneously:
nearest-neighbor CO-O and N-N pairs cannot reside on the lattice.

We now define the Markov process associated with the above set of
reactions. On a lattice ${\cal L}$ comprising $N$ sites,
the state-space of the process is the set
of configurations
$\{ \sigma \} \equiv \{ \sigma_i \}_{i \in {\cal L}}$, where the site
variable $\sigma_i$ takes values V, N, C or O in case site $i$ is vacant,
or occupied by N, CO, or O, respectively.
One trial (or sample path) of the process consists of a sequence of
configurations $\{ \sigma \}_0,...,\{ \sigma \}_{\cal N} $.
A transition between
configurations $\{ \sigma \}_n $ and $\{ \sigma \}_{n+1} $ is generated via
the following steps:
\begin{flushleft}

(1) Choose the identity of the next arriving molecule:
    CO with probability $y$, NO with probability $1-y$.
    
(2) In case of CO, choose a site {\bf x}; in case of NO, choose
    a nearest-neighbor pair ({\bf x}, {\bf y}).  If {\bf x}
    (and/or {\bf y}, in the case of NO) is occupied in $\{ \sigma\}_n$, 
    then $\{ \sigma \}_{n+1} = \{ \sigma\}_n$, i.e., the configuration
    does not change.  Otherwise, let $\{ \sigma '\}$ be the configuration
    obtained by placing CO at {\bf x}, or in the NO case, N at {\bf x}
    and O at {\bf y}.

(3) If $\{ \sigma '\}$ contains no N-N or CO-O nearest-neighbor pairs,
    then $\{ \sigma \}_{n+1} = \{ \sigma ' \}$.  
    If $\{ \sigma '\}$ does contain such pairs, they will react.
    Specifically, if the newly-arrived CO has $m$ neighbors in state O,
    then one of these (chosen at random if $m>1$), as well as the CO,
    is removed from $\{ \sigma '\}$ to give $\{ \sigma \}_{n+1}.$
    In the case of NO deposition, the analogous procedure is applied to
    the newly-arrived N atom (if it has one or more neighbors N), and to
    the O atom (should it have any CO neighbors),
    to generate $\{ \sigma \}_{n+1}.$
\end{flushleft}

We associate with configuration $\{ \sigma \}_n $ a ``time" $t = n/N$.
(This adds nothing to the process; it is convenient, nonetheless, to define
a time unit comprising one attempted transition, on average,
per lattice site.
In simulations, it is often more
efficient to choose the {\it first} site {\bf x} for the adsorption
step from a list of currently vacant sites.  But in this case, the
time increment associated with the transition from  
$\{ \sigma \}_n $ to $\{ \sigma \}_{n+1} $ is $\Delta t = 1/{\cal V}_n$
where ${\cal V}_n$ is the number of vacant sites in $\{ \sigma \}_n $.)

To our knowledge, all of the simulation studies of the NO+CO cited
herein treat the Markov process defined above.  (The constant-coverage
studies of BZ clearly follow a different procedure, but the stationary
properties of the two processes should converge in the
large-size limit \cite{bz}.)  In the case of diffusion, however,
various definitions have been employed.  Here, we implement diffusion
of (for example, N atoms) by modifying steps (1) --- (3) as follows.
Prior to (1), we impose
\begin{flushleft}
(0) Choose the process: diffusion with probability $D/(1+D)$, adsorption
    with probability $1/(1+D)$.  In the latter case, proceed to steps
    (1) --- (3) as above.  In the former, perform instead:

(1') Choose a site {\bf x} at random.  If {\bf x} is {\it not} occupied by
    N, then $\{ \sigma \}_{n+1} = \{ \sigma\}_n$.  Otherwise, choose a
    neighbor {\bf y} of {\bf x} at random.  If {\bf y} is occupied, 
    $\{ \sigma \}_{n+1} = \{ \sigma\}_n$.  Otherwise, let
    $\{ \sigma ' \}$ be $ \{ \sigma\}_n$ with {\bf x} vacant and {\bf y}
    occupied by N.

(2') If $\{ \sigma ' \}$ is free of N-N pairs,
     $\{ \sigma \}_{n+1} = \{ \sigma ' \}$.  Otherwise, choose a
     reacting pair, as in step (3) above, to generate $\{ \sigma \}_{n+1}$.
\end{flushleft}

\noindent This procedure is not calculated to optimize computational
efficiency, but rather to provide a meaning for the parameter $D$ that
is {\it independent of the configuration}. The rate of
hopping attempts of an adsorbed N atom is $D/(1+D)$.
Diffusion processes for other species are defined analogously.

\section{Pair Approximation}

Before considering the NO+CO model in detail, we explain a
simplified method for deriving the PA equations.
These equations govern the evolution of the probabilities
$P(ij,t)$, that a randomly chosen nearest-neighbor pair of
sites, {\bf x} and {\bf x}', say, are in states $i$ and $j$.
In most previous treatments, (see, for example Ref. \cite{rd86}),
the rates of change of the $P(ij)$ are derived by enumerating the
changes in the number of nearest-neighbor pairs in a {\it neighborhood} of
sites centered on, and including, a central pair.
For example, in a one-dimensional system with
nearest-neighbor interactions and two states, `0' and
`1', per site, a transition of the form $(0011) \rightarrow (0111) $
occurs at rate $P(0011) w(001 \rightarrow 011)$.
Counting the changes in the central pair {\it and} the periphery,
we see that in this process,
one (11) pair is created and one (00) pair destroyed.  The simplification
comes from {\it ignoring} changes outside the central pair, and regarding
the above process as one in which $(01) \rightarrow (11)$.
Since $P(ij)$ is the probability for {\it any} nearest-neighbor pair,
following the
changes at a particular pair (e.g., the central one), is sufficient.
This results in a significant reduction in bookkeeping, particularly
in two or more dimensions, and for processes (such as the NO+CO model)
in which a fairly large number of peripheral sites can influence
the transition probabilities.  We illustrate the method by applying it
to the contact process in Appendix A.

In the table below we list the allowed ($\surd$) and forbidden
($\oslash$) nearest-neighbor pairs in the NO+CO model.
(Entries below the main diagonal are redundant.)  

\[
\begin{array}{|c|c|c|c|c|}
\hline
         & \mbox{V} & \mbox{N} & \mbox{C} & \mbox{O} \\
\hline
\mbox{V} & \surd  & \surd    & \surd & \surd   \\
\hline
\mbox{N} &        & \oslash  & \surd & \surd    \\
\hline
\mbox{C} &        &          & \surd & \oslash  \\
\hline
\mbox{O} &        &          &       & \surd   \\
\hline
\end{array}
\]

Next we require the set of transitions between pairs.
In the table below,
we assign arbitrary labels to the allowed transitions,
and leave the remaining fields blank.  For the
square lattice only processes 1 - 20 are pertinent, 21 and 22 being
possible only on the triangular lattice.  In one dimension, transition
4 is also excluded.  Diffusion alters the rates, but not the set of
possible transitions, the only exception being when we consider 
sublattices.

\[
\begin{array}{|r|c|c|c|c|c|c|c|c|}
\hline
\mbox{From:} & \mbox{VV}& \mbox{VN}& \mbox{VC}& \mbox{VO}
  & \mbox{NC} & \mbox{NO} & \mbox{CC} & \mbox{OO} \\
\hline
\mbox{To: VV} &    & 1  & 2  & 3  & 4 &    &    &    \\
\hline
\mbox{VN}     & 5  &    & 21 &    & 6 & 7  &    &    \\
\hline
\mbox{VC}     & 8  &    &    &    & 9 &    & 10 &    \\
\hline
\mbox{VO}     & 11 & 22 &    &    &   & 12 &    & 13 \\
\hline
\mbox{NC}     &    & 14 & 15 &    &   &    &    &    \\
\hline
\mbox{NO}     & 16 & 17 &    & 18 &   &    &    &    \\
\hline
\mbox{CC}     &    &    & 19 &    &   &    &    &    \\ 
\hline
\mbox{OO}     &    &    &    & 20 &   &    &    &    \\
\hline
\end{array}
\]

The heart of the calculation lies in deriving expressions for the transition
rates $R_1,...,R_{22}$.  Once these are in hand,
we can write the equations for the pair probabilities by noting
that each transition acts as a source for one pair (i.e., in the
first column of the table), and a loss term for another (listed in
the top row).  Denoting the pair probabilities by $(ij)$, where $i$ and $j$
can be $V$, $N$, $C$, or $O$, we have, for example, that

\begin{equation}
\frac{d (VN)}{dt} = R_5 + R_6 + R_7 + R_{21} - R_1 -R_{14} - R_{17} -R_{22},
\end{equation}
and

\begin{equation}
\frac{d (VV)}{dt} = 2[R_1 +R_2 +R_3 +R_4 - R_5 -R_8 -R_{11} -R_{16}],
\end{equation}
the overall factor of two arising, in the last expression, due to 
contributions in which the molecules (always of different species, in this case)
occur in the opposite order.  Here it is important to emphasize that
for $i \neq j$, $(ij)$ represents the probability to find a site {\bf x} in
state $i$, and its neighbor {\bf y} in state $j$; $(ji)$, which represents the
reversed situation, is of course equal to $(ij)$ by symmetry.  Thus
the normalization condition reads

\begin{equation}
{\mbox (VV) + (CC) +(OO) + 2[(VN) +(VC)+(VO) +(NC) +(NO)]} = 1.
\label{norm}
\end{equation}

We introduce a similar notation for site probabilities:
\begin{equation}
(V) = (VV) + (VN) + (VC) + (VO) ,
\end{equation}
\begin{equation}
(N) = (NV) + (NC) + (NO) ,
\end{equation}
\begin{equation}
(C) = (CV) + (CN) + (CC)  ,
\end{equation}
and
\begin{equation}
(O) = (OV) + (ON) + (OO) .
\end{equation}
Another useful piece of notation represents the probability of having site {\bf x}
in state $i$ and its neighbor {\it not} in state $j$ by $(i \!\!\not \! j)$.  For example,

\begin{equation}
(V \!\!\not \!\!N) = (VV) + (VO) + (VC) ,
\end{equation}
and
\begin{equation}
(C\! \! \not \!C) = (CV) + (CN).
\end{equation}

Finally, when employing sublattices $A$ and $B$, we use
$(i)_A$ to denote the site probability of $i$ in sublattice $A$
(similarly for $(i)_B$), and
$(ij)_A$ to denote the probability of finding a site in state $i$ in the $A$
sublattice, and its neighbor (in $B$), in state $j$.
$(i \!\! \not \! j)_A$ is defined analogously.  In a sublattice
calculation we have thirteen different pair probabilities, since for $i \neq j$
we must distinguish $(ij)_A$ and $(ij)_B$.  (By definition, $(ij)_A = (ji)_B$.)
Diffusion introduces one further transition beyond those enumerated above: if species $i $ can perform nearest-neighbor hopping,
the transition $(iV)_A \rightarrow (iV)_B $ becomes possible. 

Each (nondiffusive) event involves the arrival of a molecule, either CO or NO, at the surface.
Since the next arriving molecule is CO with probability $y$,
rates for processes involving the arrival of CO carry a factor of $y$.
In processes involving the arrival of NO, we require
the probability that the next event involves N arriving at a certain site
{\bf x}, and, of course,
O arriving at a neighbor, {\bf y}.  In a lattice with coordination number $z$, this probability is 
$\tilde{y} \equiv (1-y)/z$.

The expressions for the various rates are, in general, quite complicated, and we shall not
list all of them here.  Examples of their derivation are given in
Appendix B \cite{offer}.  The PA equations are integrated numerically,
using a fourth-order Runge-Kutta scheme \cite{rec},
starting from an empty lattice.

\section{Results}

\subsection{One dimension}

Applied to the NO+CO
model on a line, the PA
predicts no active steady state in the absence of diffusion (see Fig. 1). 
The dependence of the coverages on $y$ is qualitatively similar to that
found in simulations of the square lattice \cite{mwe,yk94}.
The vacancy fraction is nonzero only for $y=0$, where we find 
$(V)=0.1623$, $(O)=0.5013$, and $(N)=0.3364$ in the stationary state.
In fact, the final vacancy concentration should be the
same as in one-dimensional dimer RSA (without reaction), i.e.,
$(V)= e^{-2}$ = 0.13534... \cite{flory}.
This is because a N-N reaction always yields the configuration
OVVO.  The vacancy pair is subsequently filled in, so sum of the
final coverages, $(N) + (O)$, is identical to the final O coverage
in RSA of O$_2$ \cite{knote}.
While the PA is exact for dimer RSA in one-dimension
(due to a shielding property \cite{evans}), in the present case
the approximation cannot deal adequately with the reactions.
The configuration VNV, for example, is impossible in the pure-NO
process (each N has at least one O neighbor), but is assigned a
nonzero probability in the PA.  We obtain better results
for the final coverages from a three-site approximation:
$(V)=0.1486$, $(O)=0.5001$, and $(N)=0.3505$.
(The KKN method yields a further slight improvement:
$(V)=0.1453$, $(O)=0.5095$, $(N)=0.3452$; simulations yield
$(V)=0.1353$, $(O)=0.5066$, and $(N)=0.3581$ \cite{kup}.)

Next we consider the effect of a nonzero diffusion rate, $D_N$,
of N$_a$ atoms.   For $D_N > D_N^c = 4.38$,
we find a reactive window for $y_1 < y < y_2$,
with a continuous (discontinuous) transition at $y_1$ ($y_2$).
Fig. 2 shows the coverages for
$D_N=10.0$. In Fig. 3 we plot $\Delta \equiv y_2 - y_1$ as a
function of $D_N$. 
With increasing $D_N$,
$y_1$ tends to zero and $y_2$ to $0.2$.
Close to $D^{c}_{N}$, $\Delta$ is described, approximately,
by $\Delta \sim (D_N - D^{c}_{N})^{0.67}$.
For diffusion of CO (but none of the other species), we find an active 
state for $D_{CO}\geq D^{c}_{CO}=7.19$. The width, which is
very small, grows linearly with 
$D_{CO}$ in the neighborhood of the critical value. 
Both $y_1$ and $y_2$ shift to higher values with increasing $D_{CO}$.
Finally, for diffusion of $O$ atoms (exclusively), we find 
$D^{c}_{O}=0.75$.  The window width follows 
$\Delta \approx (D_O - D^{c}_{O})^{0.24}$ in the vicinity
of $D_O^c$. The continuous transition ($y=y_1$) always occurs
near $y=0$; for large values of $D_O$, 
$y_2$ approaches a limiting value of 0.27.  
We note that for $y=0$, even 
very small values of $D_O$ cause a drastic reduction in the 
fraction of vacant sites: 
when $D_O=0.05$, for example, (V)$<10^{-4}$; for
$D_O=16.0$, (V) $<10^{-6}$.  Similar behavior is observed in
simulations on the triangular lattice \cite{kn}.

\subsection{Triangular lattice}

The coverages predicted by our method
for the NO+CO model (without diffusion) on the triangular lattice
are compared against simulation results \cite{mwe} in Fig. 4.
There is an active steady state between
the continuous transition at $y_1=0.040(1)$, and $y_2=0.363(1)$,
which marks a discontinuous transition.  (We note that the latter
is the result for an initially empty lattice.  Commencing with
a finite CO-coverage will in general yield a smaller value for $y_2$.)
The table below compares our PA results for the transition
points with those predicted by the site approximation (SA) \cite{bz,mwe},
and the KKN method \cite{kkn}, and found in simulations \cite{bz}.

\[
\begin{array}{|c|c|c|c|c|}
\hline
         & \mbox{SA} & \mbox{ KKN}  & \mbox{PA} & \mbox{Simulation} \\
\hline
\mbox{$y_1$} &  0.0      & 0.152(1)   & 0.040 & 0.1725(25)    \\
\hline
\mbox{$y_2$} &  0.3877   & 0.393(1)   & 0.363 & 0.3514(1)  \\
\hline
\end{array}
\]
While our result for $y_2$ is in good agreement with simulation, we obtain
a poor estimate for $y_1$. The latter is a consequence of 
neglecting explicit correlations beyond nearest neighbors 
in the pair approximation, and, perhaps, of ignoring certain
nearest-neighbor pair factors in reckoning cluster
probabilities (see Appendix B).
For $y=0$ we find a final poisoned state characterized by 
$(O) = 0.8538$, $ (N) = 0.1462$, and $(V) \simeq 2 \times 10^{-6}$. 
Simulations yield 
$(O) = 0.7443$, $(N) = 0.1656$, and $(V) = 0.0901$ \cite{ziffup}.
Clearly the PA does not give an accurate description of this
RSA process with partial reaction, for either the triangular
or square lattices (see below).  (This is in contrast to pure
dimer deposition, for which the PA does reasonably well \cite{rd86}.)
Despite these discrepancies, the PA coverages are generally in good
agreement with simulation.

\subsection{Square lattice}

The PA prediction for the phase diagram
is qualitatively similar to that found for the triangular lattice.
The continuous transition from a predominantly O-poisoned state
to an active state occurs at $y_1 = 0.111$, and the discontinuous
transition falls at $y_2 = 0.2981$.  (The latter, again, is determined
using an initially empty lattice.)  Cort\'es et al. obtained
$y_1 \approx 0.09$ and $y_2 \approx 0.35$ in this case, showing that
the large-reaction-rate limit of a calculation using finite rates
yields results comparable, but not identical to, one employing
instantaneous reactions.  (While the comparison is academic in the
present instance, there being no active state on the square lattice,
it is of interest to gauge the agreement between the two
methods.)  For $y=0$ we obtain $(O) = 0.6551$, $(N) = 0.0198$,
while simulations yield
$(O) = 0.6495$ and $(N) = 0.2416$ \cite{ziffup}.

Since the PA predicts no active state in one dimension
(in the absence of diffusion), it is interesting to check whether
removing reaction 4, which is impossible in one dimension, changes
the phase diagram.  We find that deleting this process has a minimal
effect on the PA prediction for the square lattice.  The cause for
the dramatic difference between the linear and square lattice
results must be sought elsewhere.  (It is worth remarking that the PA
similarly yields only absorbing states for the one-dimensional
ZGB model.)

Introducing sublattices in the PA calculation
(yielding what we call the PAS theory), changes the result drastically.
We find that the active state predicted by the PA is
{\it unstable} to the formation of N-rich and N-poor sublattices;
in the process, the vacancy density falls to zero, and the
active state vanishes.
As detailed in Appendix B, we employ an extended set of
variables, i.e., for $i \neq j$, ($i,j = V, N, C$, or $O$),
probabilities $(ij)_A$ and $(ij)_B$, representing 
species $i$ in the $A$ or the $B$ sublattice.
(Naturally, the site coverages $(N)_A$ and $(N)_B$, etc., may
also differ.)
We begin the calculation, as before, with an empty lattice;
the pair equations
reach the same steady solutions as in the simple PA.  For
$y_1 < y < y_2$, we then probe the stability of the active state
by transferring a small amount,
$\Delta (N) \equiv (N)_A - (N)_B = 10^{-3}$,
of N from one sublattice to the other, and study the response to
this small perturbation.  We find, in all cases, that $\Delta (N)$
grows, and that $(V)$ decreases, finally becoming zero, that is, the
PAS equations reach an absorbing state.
Essentially identical results are obtained if we start from a slightly
asymmetric initial condition, i.e., with a small N-coverage on one of the
sublattices, as was done in Ref. \cite{kkn}.
(We note in passing that introducing sublattices in the {\it site}
approximation has no effect on the results.)

The PAS, then, represents the first theoretical approach giving
a phase diagram in accord with simulation, for the NO+CO model on
the square lattice.  But it retains some of the undesirable features
of the PA.  Fig. 5 shows that for $y < 0.1$, and again for $y > 0.5$,
the PAS coverages are in good agreement with simulation.  In these
regimes, of course, the PA and PAS are identical (they only differ
on the interval where the PA predicts an active state).
For $0.1 < y < 0.5$ there are substantial differences between theory
and simulation, associated with the continued appearance of phase
transitions at $y_1$ and $y_2$ in the PAS.  For $y_1 < y < y_2$ the
PAS equations exhibit spontaneous symmetry breaking \cite{kkn}
associated with global ``antiferromagnetic" order, $\Delta (N) \neq 0$,
(see Fig. 6), which, as we have remarked, is not seen in simulations.

A final point concerns the effect of N diffusion on the phase diagram.
Since this process tends to equalize the sublattice coverages, it is
reasonable to expect an active state for sufficiently large values of
$D_N$.  This is indeed observed in simulations \cite{yk94,kn98},
where the range $\Delta$ of $y$ values supporting an active state
grows steadily with $D_N$.  (The PA, as we have noted,
yields a similar result in one dimension.)  When we include N diffusion
in the PAS calculation, we observe no active state for
$D_N < D_N^{c} \simeq 0.023$, but for larger diffusion rates the entire
interval between $y_1$ and $y_2$ becomes active at once.

\section{Discussion}

We have formulated a pair approximation (PA) for the NO+CO
surface reaction model with instantaneous reactions, using a simplified
derivation. 
The PA gives quite reasonable predictions for
coverages on the triangular lattice, but a surprisingly
low value for the continuous transition point, $y_1$.
The pair
approximation with sublattices (PAS) gives a qualitatively correct
phase diagram for the square lattice, but certain anomalous features of
the simple PA persist,
notably, singular behavior of the coverages at the transition points
$y_1$ and $y_2$.

Our study shows that the PA can furnish reliable qualitative,
and in some instances quantitative predictions for  
reaction models, {\it provided it includes a mechanism for realizing any
symmetry-breaking tendency inherent in the model}.  The same condition
applies, of course, to mean-field or cluster approximations for
equilibrium models.  In general, it is asking too much
of such theories to provide quantitatively reliable phase boundaries;
the PA may nonetheless yield some insight into the overall shape of
the phase diagram.  More accurate theories are typically based on the
hierarchy of $n$-point functions.

The theory of KKN yields a remarkably accurate prediction for $y_1$
on the triangular lattice.
This indicates that
the effect of long-range correlations
is reasonably well-represented in their theory.
Surprisingly, the PA yields a better
prediction for $y_2$.
It remains to develop a method that combines the
advantages of the two approaches.
\vspace{2em}

{\bf Acknowledgments}
\vspace{1em}

We are grateful to Robert M. Ziff and Olaf Kortl\"uke
for very helpful correspondence,
and for providing us with simulation data prior to their publication.
We thank Jim Evans for valuable correspondence.
A.G.D. and W.F. are supported by CNPq (Brazil).  We thank
FINEP for financial support.

\newpage
{\bf Appendix A: Simplified derivation of pair equations}
\vspace{1em}

Here we illustrate our method by applying it to the contact
process (CP) \cite{harris}.  The CP is a Markov process defined on a
$d$-dimensional cubic lattice.  Each site is
either vacant (0) or occupied (1).  The
transition rates at any site are $w(1 \rightarrow 0) = 1$ (independent of
the neighbors) and $w(0 \rightarrow 1) = \lambda n /2d$, where
$0 \leq n \leq 2d$ is the number of neighboring sites in state 1.
(Note that all sites 0 is an absorbing configuration.)
Thus in one dimension we have $w(101 \rightarrow 111) = \lambda$,
$w(100 \rightarrow 110) = w(001 \rightarrow 011) = \lambda/2$, and
$w(000 \rightarrow 010) = 0.$  We enumerate below the transitions,
associated rates (e.g., the transition rate times the
probability of the initial configuration),
and overall changes in the number of 1's and 11
nearest-neighbor pairs, in the one-dimensional CP.
We use (1) to denote the density of 1's, (10) for the probability of
a nearest-neighbor 0---1 pair, etc.  (By normalization,
$(00) + 2(01) + (11) = 1$.  Note that the second and fourth
entries carry a factor of 2 to account for mirror-image
events.)

\[
\begin{array}{|c|c|c|c|}
\hline
\mbox{Process} & \mbox{Rate} & \Delta N_{1} & \Delta N_{11} \\ 
\hline
101 \rightarrow 111 & \lambda (01)^{2} /(0) & +1  & +2  \\ 
\hline
100 \rightarrow 110 & 2\lambda (01)(00)/(0) & +1  & +1  \\ 
\hline
111 \rightarrow 101 &     (11)^{2} /(1)     & -1  & -2  \\ 
\hline
110 \rightarrow 100 &     2 (11)(01)/(1)    & -1  & -1  \\ 
\hline
010 \rightarrow 000 &     (01)^2/(1)        & -1  & 0   \\
\hline
\end{array}
\]

\noindent Collecting results, one finds

\begin{equation}
\label{pmftrho}
\frac {d(1)} {dt} = \lambda [(1) \!- \! (11)] - \; (1),
\end{equation}

\noindent and

\begin{equation}
\label{pmftz}
\frac {d(11)} {dt} = \lambda \frac {(1) \! - \! (11)} {(0)} 
 [1 - (11)] \; - \; 2(11).
\end{equation}

In this approximation, an active stationary solution (one with $(1) > 0$),
exists only for $\lambda > \lambda_c = 2$.  The above calculation is
simple enough in one dimension, but becomes more complicated for
higher $d$.  We illustrate our simpler alternative method below
on the $d$-dimensional CP.  Only transitions at the central pair
are enumerated.  The rates involving creation receive
independent contributions
from within the pair (if it is of type 01) {\it and} from the $2d-1$
neighbors outside.  (This independence is of course an approximation
intrinsic to the PA.)

\[
\begin{array}{|c|l|}
\hline
\mbox{Transition} & \;\;\;\;\;\;\;\;\;\;\;\;\;\;\;\;\;\;\; \mbox{Rate} \\ 
\hline
11 \rightarrow 01 & R_1 = (11)  \\ 
\hline
01 \rightarrow 11 & R_2 = \lambda (01) \left[1 + (2d-1) (01)/(0) \right]/2d \\ 
\hline
10 \rightarrow 00 & R_3 = (10)  \\ 
\hline
00 \rightarrow 10 & R_4 = (2d-1)\lambda (00)(01)/2d(0) \\  
\hline
\end{array}
\]

\noindent We find $d(11)/dt = 2[R_2 - R_1]$, which
reduces to Eq. (\ref{pmftz}) for $d=1$, when we note that
(01) = (1) - (11).
Noting that (1) = (11) + (10),
we immediately recover Eq. (\ref{pmftrho}) for any $d$.
(Analysis of the stationary solutions shows that 
$\lambda_c = 2d/(2d-1)$ in $d$ dimensions.)
\vspace{2em}

{\bf Appendix B: Rates for the NO+CO model}
\vspace{1em}

In this Appendix we present several examples of the
pair approximation (PA) transition rates
for the NO+CO model, first on the square, and then
on the triangular lattice.  
We begin with transition 15, VC $\rightarrow$ NC, on the
square lattice.  Fig. B1 shows one of three equivalent
configurations needed to realize this process.
The rate (per bond of the lattice) carries the factors
$(VC)$ (probability of the initial state) and $\tilde{y}$ (probability
of NO arriving with N falling at the central vacant site).
There are three possible locations (neighbors of V) at
which O might adsorb; each is vacant, in the PA, with probability
$(VV)/(V)$.
(Whether the O atom reacts or not is unimportant in this instance.)
If either of the remaining neighbors of V harbors an N atom,
the newly-arrived N will react.  (Recall that we are considering
{\it infinite} reaction rates in this work.)  Hence these two sites
must be free of N for the desired transition to occur, implying a
factor of $[(V \!\! \not \!\!N)/(V)]^2$.  Thus

\begin{equation}
R_{15} = 3 \tilde{y} \frac{(VC)(VV)(V \!\! \not \!\!N)^2}{(V)^3}  .
\end{equation}
Now suppose that N atoms can hop to neighboring sites at rate $D_N$.
If one of the neighbors of V harbors an N atom, it can hop to the
vacant site, and will remain there if the other neighbors are free
of N.  We must then add to the above expression the diffusive contribution

\begin{equation}
D_{15} = \frac{3 D_N}{4} \frac{(VC)(VN)(V \!\! \not \!\!N)^2}{(V)^3}  \;,
\end{equation}
the factor of 1/4 reflecting the four possible directions of
the attempted hopping move.  (Note that we absorb a factor of $1+D_N$
into a rescaled time variable.)

If we consider sublattices, each process splits in two.
For process 15, the rate (including diffusion) for the case in which V lies in the
A sublattice is readily seen to be
\begin{equation}
R_{15,A} = 3 \tilde{y} \frac{(VC)_A(VV)(V \!\! \not \!\!N)_A^2}{(V)_A^3}
           + \frac{3 D_N}{4} \frac{(VC)_A(VN)_A(V \!\! \not \!\!N)_A^2}{(V)_A^3} .
\end{equation}
(For any $i$, $R_{i,B}$ is found by interchanging A's and B's in $R_{i,A}$.)

A somewhat more complicated transition is number 4, NC $\rightarrow$ VV.  It is contingent
upon an NO landing parallel to the central NC pair, with the two N's adjacent.  One of two
equivalent initial configurations is shown in Fig. B2.  The solid lines indicate the pairs
included in reckoning the probability of the configuration.  Notice
that when a pair of neighboring sites, {\bf x} and {\bf y}, are both neighbors of sites in the
central pair, we use the pair factor associated with the central pair
[(VN)(VC) in this example], in preference to the factor between {\bf x}
and {\bf y} [either (VN)(VV) or (VC)(VV)].  We apply this rule in all our
calculations, to eliminate possible ambiguities.
Consider the neighbors {\bf a} and {\bf b} of the vacant site above N.
If neither of these bear N, the newly-arrived N will surely react with the
N in the central pair.  If either {\bf a} or {\bf b} (but not both) carry
an N, the probability of the desired reaction is 1/2; it is 1/3 in case
both {\bf a} and {\bf b} bear N.  The probabilities of these
events --- no N, one N, or two --- given the vacant neighbor, are
$(V \!\!\! \not \!\!\! N)^2/(V)^2$,  $2(V \!\!\! \not \!\!N)(VN)/(V)^2$,
and $(VN)^2/(V)^2$, respectively.  An analogous consideration applies to
the probability of reaction of the central-pair CO, depending upon the
presence of CO at {\bf c} and/or {\bf d}.  Combining these observations,
we arrive at

\begin{equation}
R_{4} = 2 \tilde{y} \frac{(NC)(VN)(VC)}{(N)(C)(V)^4} 
           \left[(V \!\! \not \!\!N)(V) + \frac{1}{3} (VN)^2 \right]
           \left[(V \!\! \not \!\!C)(V) + \frac{1}{3} (VC)^2 \right] \;,
\end{equation}
where we used $(V \!\! \not \!\!N) +(VN) = (V)$ and $(V \!\! \not \!\!C) +(VC) = (V)$.
For the sublattice calculation,

\begin{equation}
R_{4,A} = 2 \tilde{y} \frac{(NC)_A (VN)_B (VC)_A }{(N)_A (C)_B (V)_A^2 (V)_B^2} 
           \left[(V \!\! \not \!\!N)_B (V)_B  + \frac{1}{3} (VN)_B^2 \right]
           \left[(V \!\! \not \!\!C)_A (V)_A  + \frac{1}{3} (VC)_A^2 \right] \;.
\end{equation}
There is no diffusive contribution to this process.

While several rates have more complicated expressions, all of the
calculations in the square lattice follow the lines of those illustrated
above.  A new question of principle does arise in the triangular lattice,
where it is not possible to include all the pair factors between the
central pair of sites and their nearest-neighbors in the simple PA.
In the triangular lattice, if
{\bf x} and {\bf y} are nearest neighbors, they have two neighbors 
(sites {\bf d} and {\bf h} in Fig. B3),  {\it in common}.
In the PA, the probability of finding {\bf x}, {\bf y} and {\bf d}
in states $i$, $j$, and $k$, resp.,
may be written as $(ij)(ik)/(i)$, $(ij)(jk)/(j)$, or $(ik)(kj)/(k)$,
but {\it not} as $(ij)(ik)(jk)/(i)(j)(k)$.
(The latter family of expressions is not, in general, even normalized!)  
Our choice of which pair factors to include is shown in Fig. B3.
Note that as we sum over all the possible states of the peripheral sites,
symmetry under the interchange of
{\bf x} and {\bf y} is restored.

As an example consider process 19, VC $\rightarrow$ CC (site
{\bf x} vacant and {\bf y} occupied by CO in Fig. B3).  A CO molecule
must land at {\bf x} and remain there, which implies that sites
{\bf a}, {\bf b} and {\bf c}
must be free of O.  (Sites {\bf d} and {\bf h} have no possibility
of bearing O,
as they are neighbors of a site occupied by CO.  The states of sites
{\bf e}, {\bf f}, and {\bf g} are unimportant in this process.)
Multiplying the independent factors associated with the events enumerated
above, we find that the rate for this process is given by:
\begin{equation}
R_{19} = y(VC)\frac{(V \!\! \not \!\! O)^3}{(V)^3}
\end{equation}

A more complicated process is number 20, VO $\rightarrow$ OO ({\bf x}
vacant, {\bf y} occupied by O in Fig. B3).
An NO must fall with O at {\bf x}, and N at one of the
sites in \{{\bf a},{\bf b},{\bf c},{\bf d},{\bf h}\}.
In no case may any of the sites in \{{\bf a},{\bf b},{\bf c}\}
hold CO.  (Being neighbors of an O, 
{\bf d} and {\bf h} are surely free of CO.)
The rate is given by the expression:

\begin{equation}
R_{20} = \tilde{y}(VO) \left\{
         3{(VV)\over (V)}\biggl({(V \!\! \not \!\! C) \over(V)}\biggr)^2
         +{(VV)\over(V \!\! \not \!\! C)} \biggl({(V \!\! \not \!\! C)\over(V)}\biggr)^3
         +{(VO)\over(O)}\biggl({(V \!\! \not \!\! C)\over(V)}\biggr)^3
         \right\} \;.
\end{equation}
The first term in brackets represents N falling at {\bf a}, {\bf b}, or
{\bf c}.
The next is for N falling at {\bf d}.
The probability of {\bf d} being vacant,
given one vacant neighbor and one occupied by O, is 
$(VV)/[(VV) +(VN) +(VO)] = (VV)/(V \!\! \not \!\! C)$.
The final term represents N falling at {\bf h}.
These examples illustrate the
principles used in the calculation, the resulting expressions,
needless to say, becoming quite involved in certain cases.

\newpage

\noindent {\bf Figure Captions}
\vspace{1em}

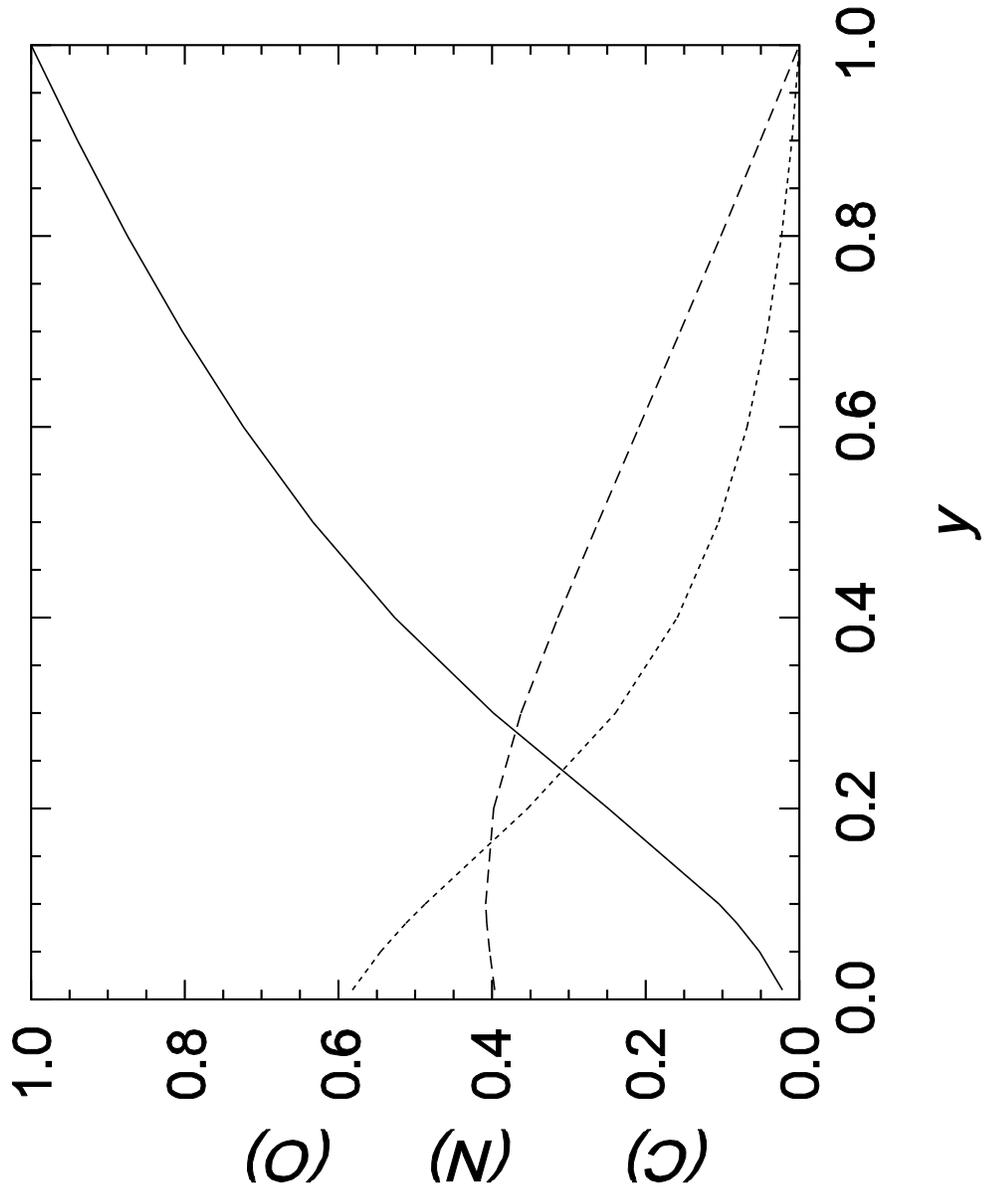
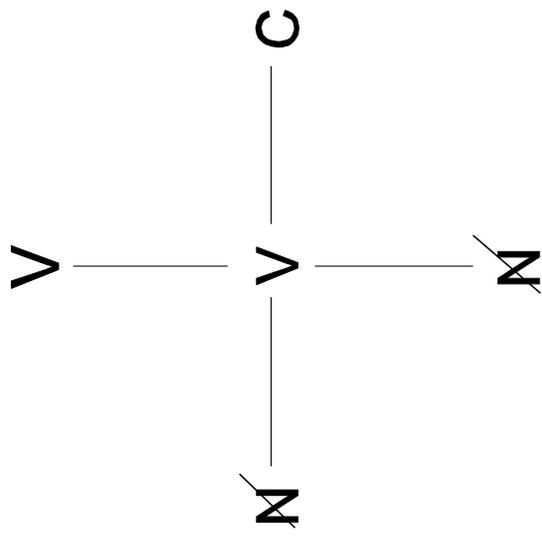
\noindent FIG. 1. Coverages versus $y$ in one dimension 
in the absence of diffusion.  Solid line: ($C$); dotted
line: $(O)$; dashed line: $(N)$.
\vspace{1em}

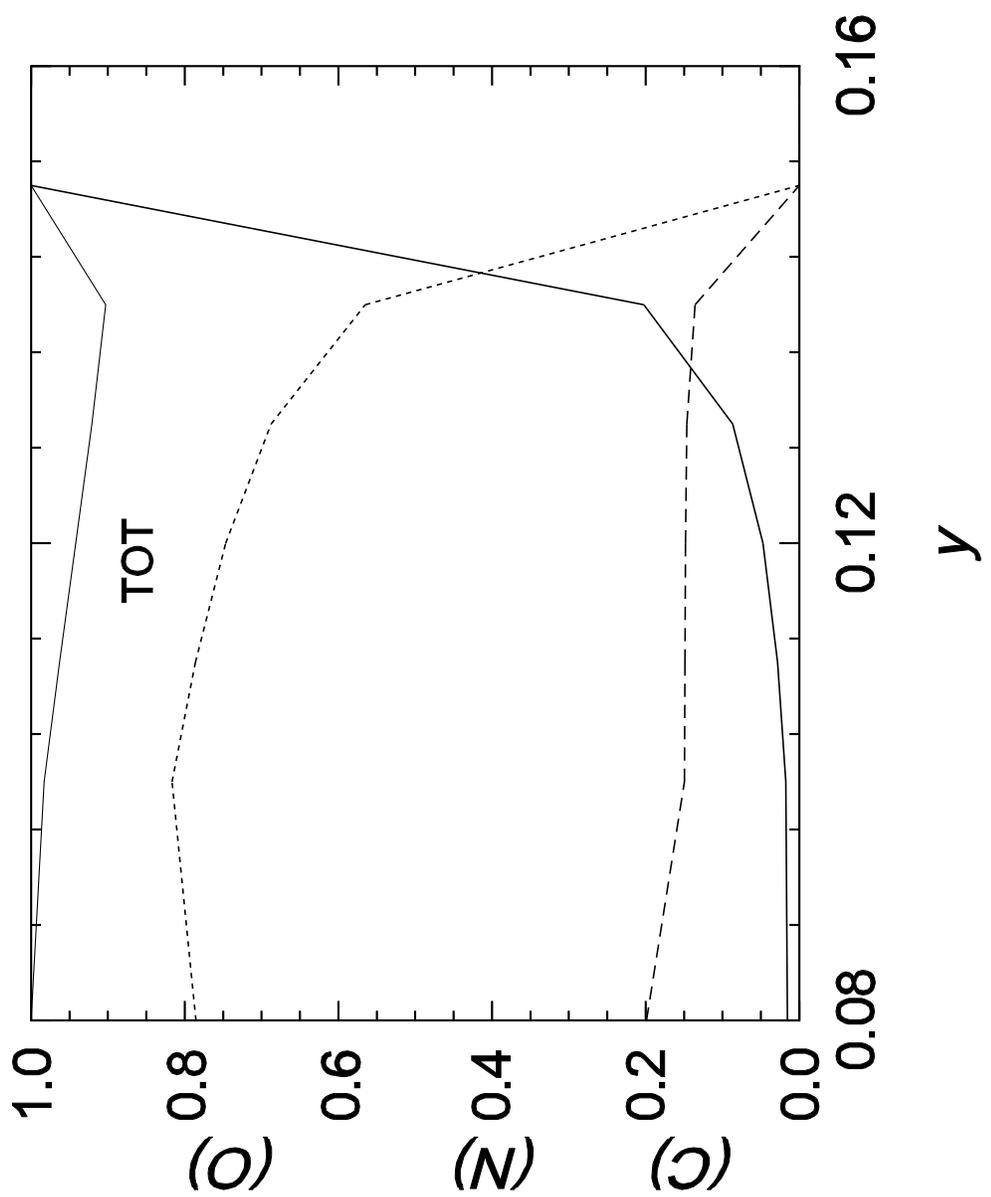
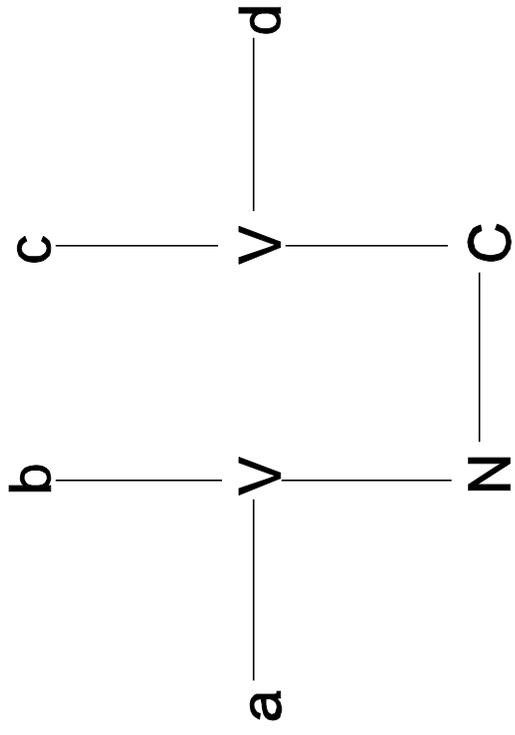
\noindent FIG. 2. Coverages versus $y$ in one dimension, for $D_N=10.0$.
Symbols as in Fig.1.
\vspace{1em}

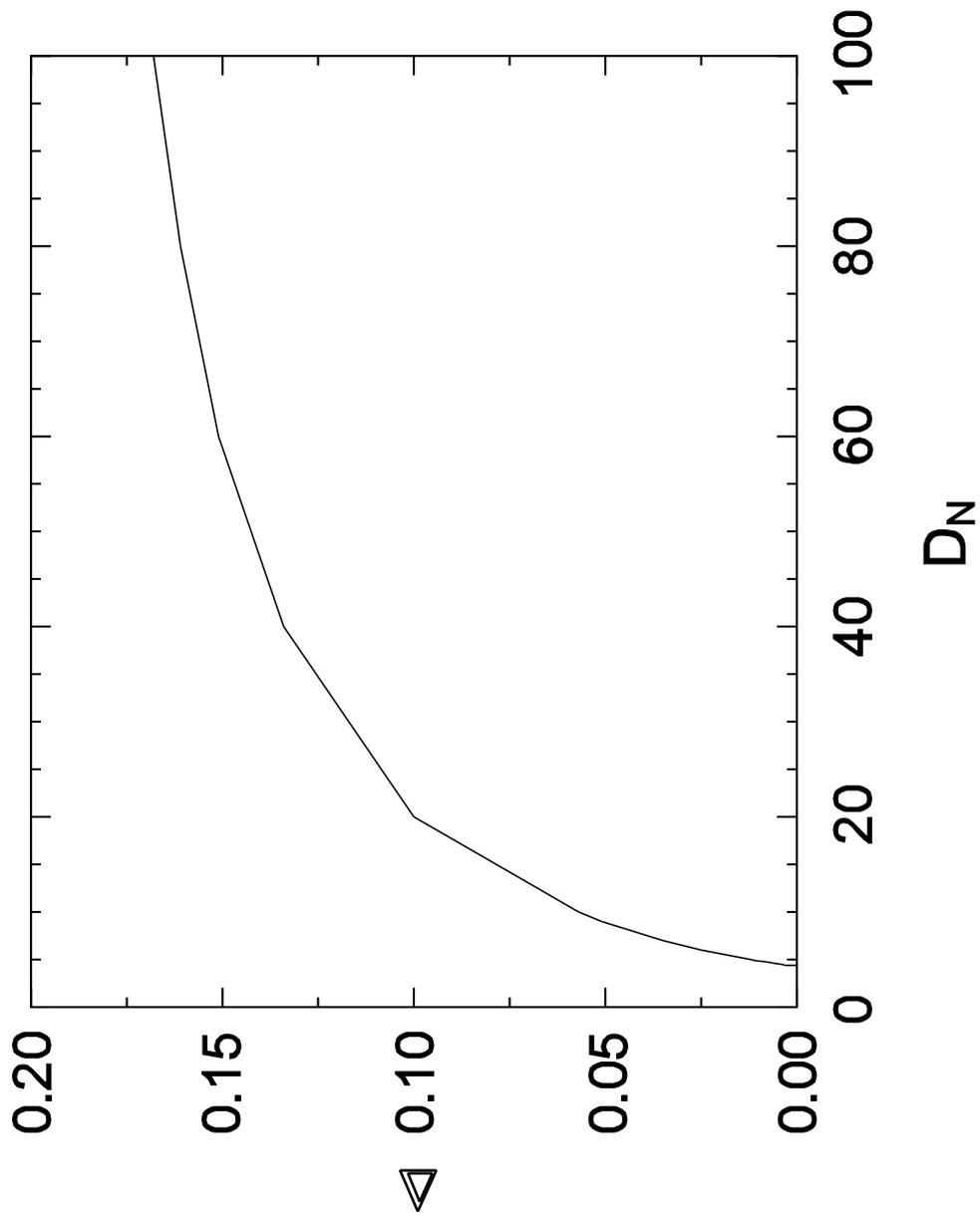
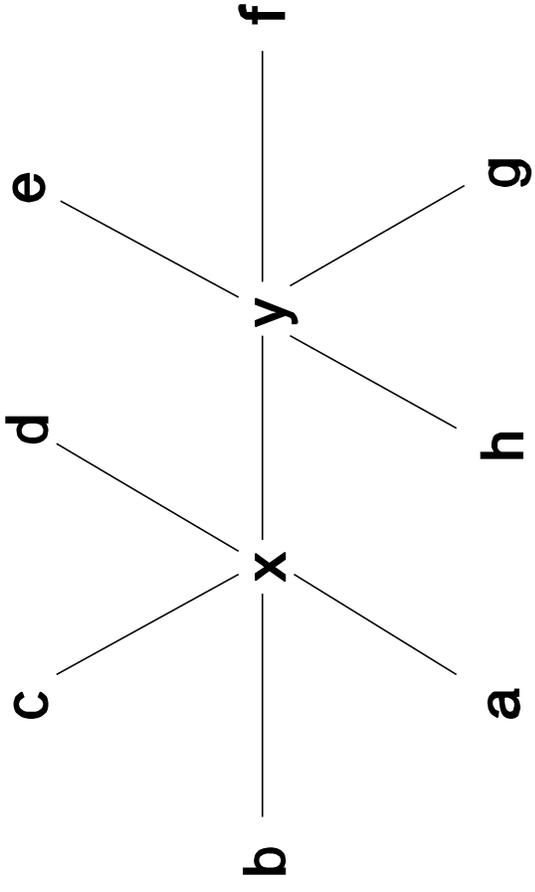
\noindent FIG. 3. Width of the reactive window in one dimension,
as a function of the diffusion rate $D_N$.
The critical value $D^{c}_N=4.38$.
\vspace{1em}

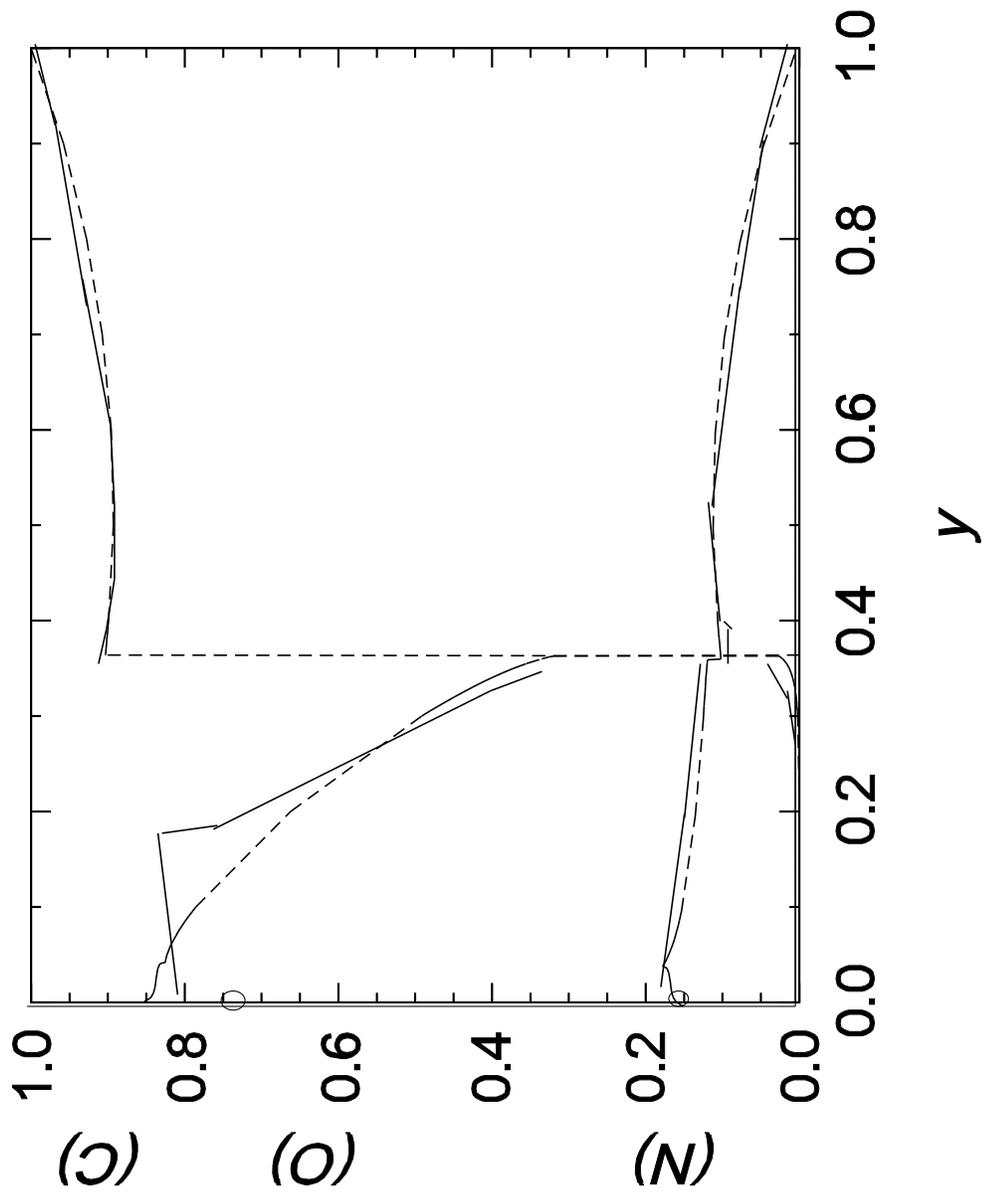
\noindent FIG. 4.  Coverages in the triangular lattice.
Solid lines: simulation (Ref. \cite{mwe}), dashed lines: PA.
\vspace{1em}

\noindent FIG. 5. a) N coverage in the square lattice.
Solid line: simulation (Ref. \cite{mwe}), dashed line: PAS.
b) CO coverage, symbols as in a);
c) O coverages, symbols as in a).
\vspace{1em}

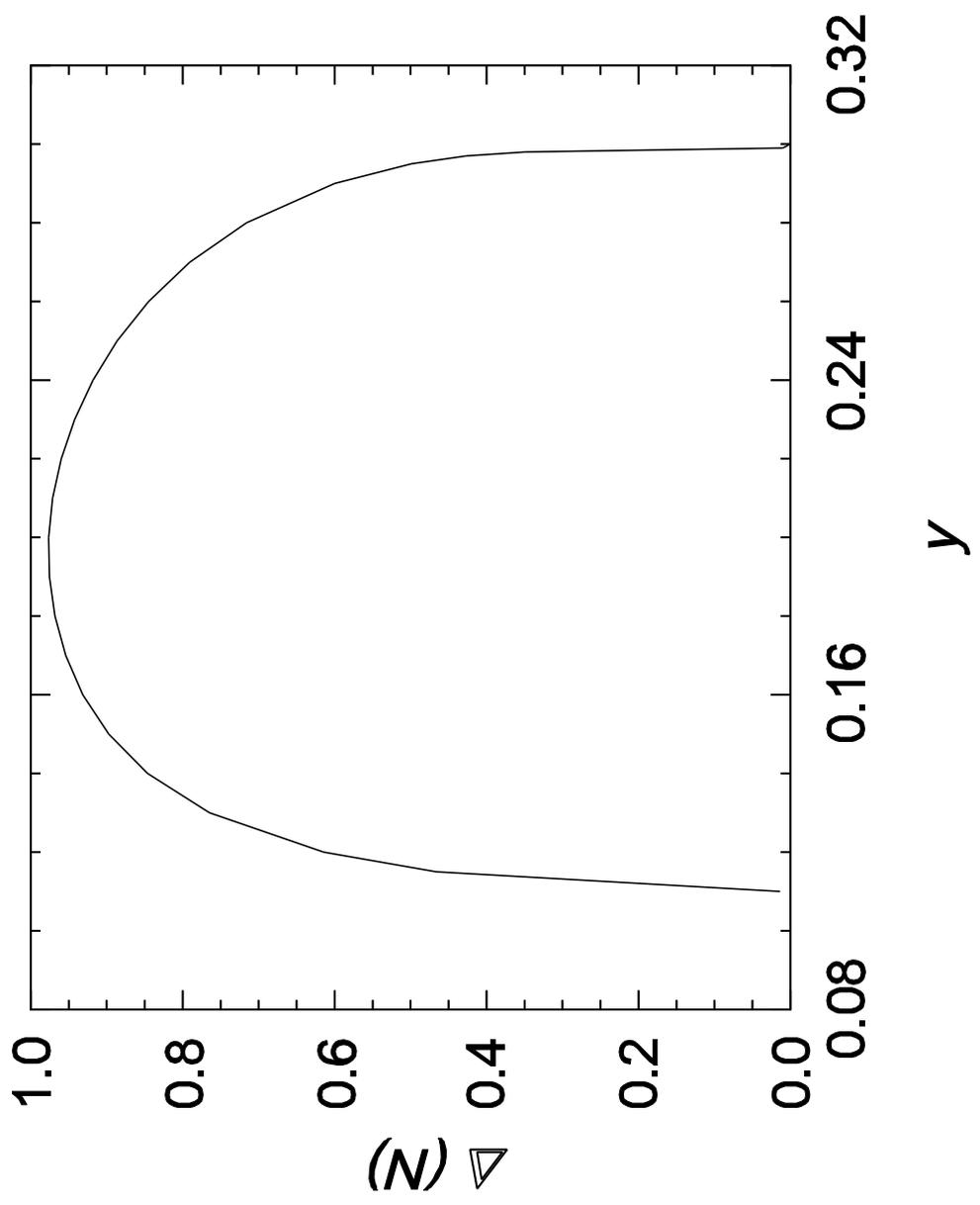
\noindent FIG. 6. PAS sublattice order parameter, $\Delta (N)$.
\vspace{1em}


\begin{thebibliography}{99}
\bibitem{offer} We shall be happy to supply detailed rate expressions and/or
        computer codes for integration of the PA equations to interested
        readers.
\end{thebibliography}
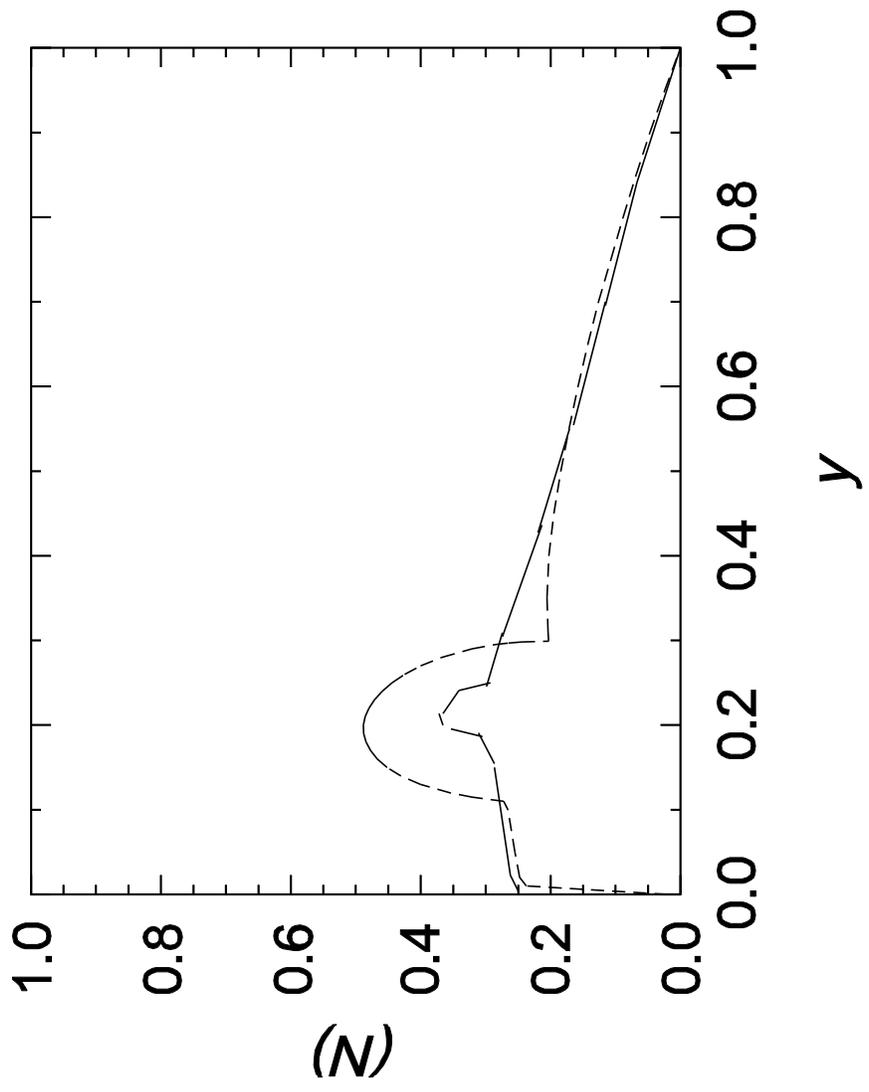
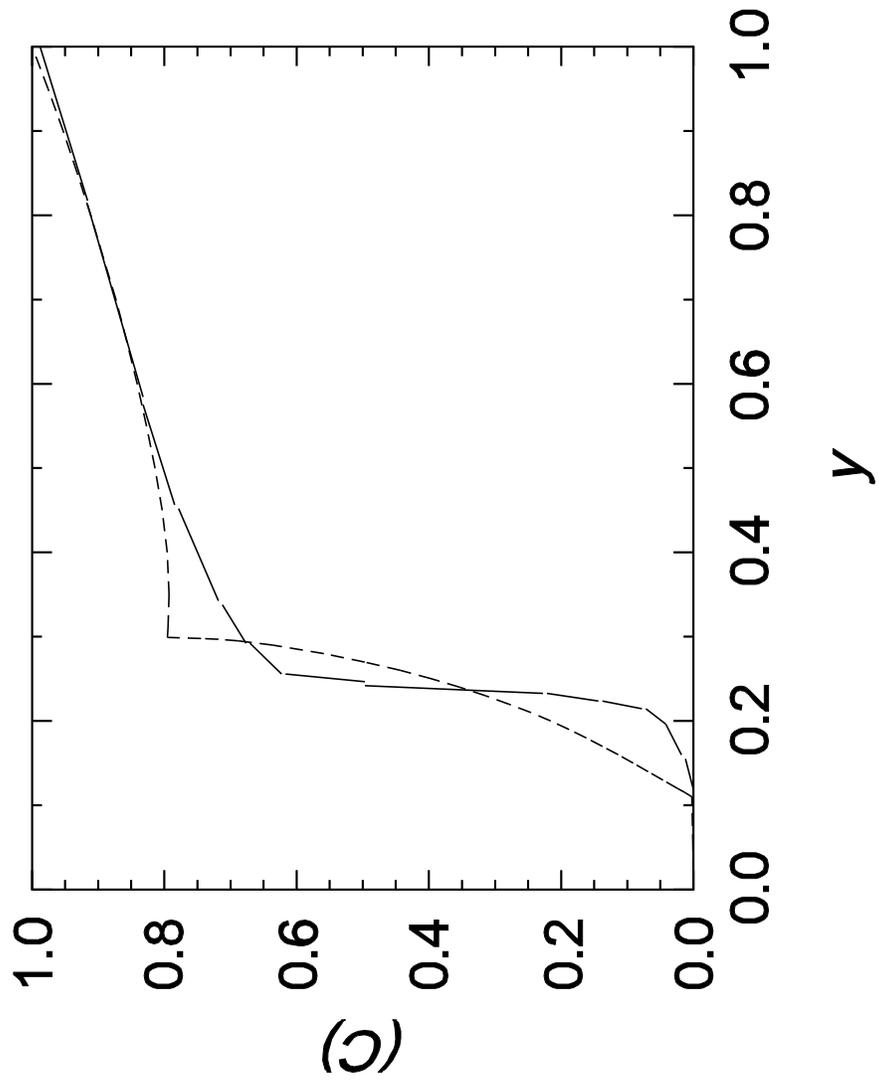
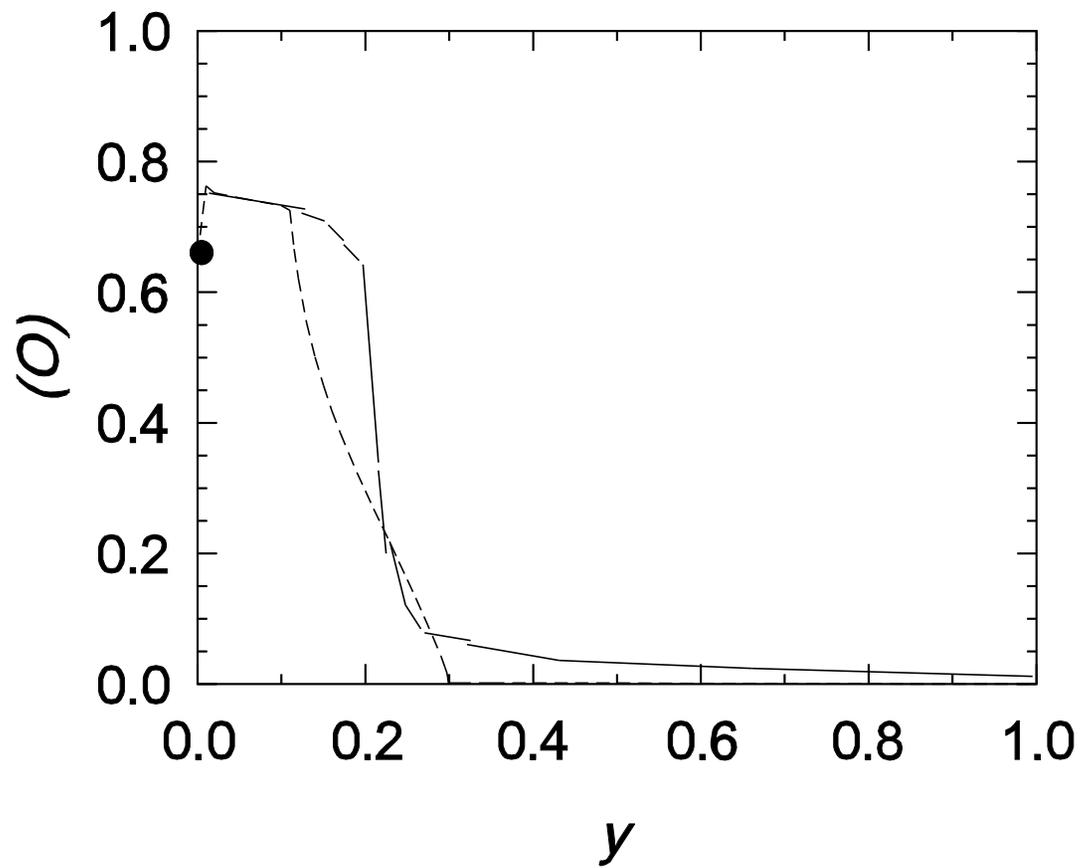
\end{document}